\documentclass[%
 aip,
 amsmath,amssymb,
 reprint,%
]{revtex4-1}

\usepackage{graphicx}% Include figure files
\usepackage{dcolumn}% Align table columns on decimal point
\usepackage{bm}% bold math
\usepackage[utf8]{inputenc}
\usepackage[T1]{fontenc}
\usepackage{mathptmx}
\usepackage{etoolbox}
\usepackage{xcolor}

\makeatletter
\def\@email#1#2{%
 \endgroup
 \patchcmd{\titleblock@produce}
  {\frontmatter@RRAPformat}
  {\frontmatter@RRAPformat{\produce@RRAP{*#1\href{mailto:#2}{#2}}}\frontmatter@RRAPformat}
  {}{}
}%
\makeatother
\begin{document}

\preprint{AIP/123-QED}

\title[Quantum and Structural effects via SACM]{Quantum and Structural Effects Captured via a Statistical Method: the SACM Applied to HCN and HNC Colliding with CO}
% Force line breaks with \\
\author{F. Tonolo}
 \email{francesca.tonolo.1@univ-rennes.fr}
 \affiliation{Univ. Rennes, CNRS, IPR (Institut de Physique de Rennes), UMR 6251, Rennes, F-35000, France}%Lines break automatically or can be forced with \\
\author{E. Quintas-Sánchez}
 \affiliation{Department of Chemistry, Missouri University of Science and Technology, Rolla, Missouri 65409, USA}%Lines break automatically or can be forced with \\
\author{A. Batista-Planas}
 \affiliation{Department of Chemistry, Missouri University of Science and Technology, Rolla, Missouri 65409, USA}%Lines break automatically or can be forced with \\
\author{R. Dawes}
 \affiliation{Department of Chemistry, Missouri University of Science and Technology, Rolla, Missouri 65409, USA}%Lines break automatically or can be forced with \\
\author{François Lique}
 \email{francois.lique@univ-rennes.fr}
 \affiliation{Univ. Rennes, CNRS, IPR (Institut de Physique de Rennes), UMR 6251, Rennes, F-35000, France}%Lines break automatically or can be forced with \\ 

\date{\today}% It is always \today, today,
             %  but any date may be explicitly specified

\begin{abstract}
This work spotlights the Statistical Adiabatic Channel Model as an efficient and accurate method for deriving low-temperature (de)-excitation rate coefficients for collisions induced by heavy projectiles. 
For such systems, fully quantum treatments become intractable, while quasi-classical methods fail at low temperature.
%For such systems, fully quantum treatments rapidly become intractable due to the large number of scattering channels, while conventional statistical and quasi-classical methods break down in the quantum-dominated low-temperature regime, limiting the predictive power of current models.
Here, we demonstrate that the Statistical Adiabatic Channel Model overcomes these limitations by combining statistical sampling with an adiabatic channel representation.
Its application to the HCN and HNC isomers colliding with CO yields rate coefficients in quantitative agreement with full quantum results benchmarked for the lowest total angular momentum. These systems are relevant for modeling cometary comae, where reliable molecular data remain scarce.
Remarkably, this approach also reproduces near-resonant energy transfer and isomeric effects, demonstrating that essential quantum and structural features can be captured within a statistical framework.
\end{abstract}

\maketitle

%\section{\label{sec:main} Main Text}

Recent advances of the new generation of telescopes are providing an unprecedented volume of accurate observational data, addressing astrophysical environments ranging from the interstellar medium to the atmospheres of comets, planets, and exoplanets \cite{gardner2006james,wootten2009atacama,dewdney2009square,pascale2018ariel}. Turning this observational richness into quantitative physico-chemical information requires models that can reliably translate line intensities into molecular abundances and temperatures. Central ingredients for this interpretation are the collisional (de)-excitation rate coefficients for the dominant projectiles. Such rate coefficients govern molecular level populations when the local thermodynamic equilibrium (LTE) approximation breaks down, $e.g.$, in the low-density regimes typical of many astrophysical environments \cite{lopez2001non,lique2009importance,bowesman2026tiramisu}. Despite their fundamental role, reliable collisional data remain scarce and often constitute a major bottleneck for the interpretation of modern observations \cite{tonolo2025collisionalb}. This is the case in environments where the dominant projectiles are heavy species: their dense rotational structure dramatically increases the number of accessible scattering channels, making fully quantum calculations computationally prohibitive in most cases. Cometary comae, the gaseous envelopes formed by sublimation of ices from cometary nuclei under solar radiation, constitute illustrative examples in this framework. They represent key tracers of astrochemical evolution because they contain the chemical signatures of the early Solar System \cite{mumma2011chemical}. However, they are dominated by CO, CO$_2$ and H$_2$O, molecular gases with small rotational constants, which significantly increases the computational cost of accurate dynamics calculations. Yet, their physical conditions (very low temperatures and densities) remain 
challenging to fully reproduce and control in laboratory experiments, despite improved experimental techniques, based on molecular beams, flow-based methods and laser spectroscopy, have now begun to provide very detailed information on collisions, including the observation of quantum resonances and interference effects 
\cite{chefdeville2012appearance,brouard2014taming,bergeat2015quantum,de2020imaging,toscano2020cold,labiad2022absolute}.
As a consequence, the interpretation of cometary observations mainly relies on accurate theoretical collisional rate coefficients. 

In this context, the Statistical Adiabatic Channel Model (SACM) has emerged as a promising alternative to fully quantum approaches. Conceptualized by Quack and Troe in the mid-1970s \cite{quack1975complex} and further developed by Loreau \emph{et al.} \cite{loreau2018efficient}, the SACM combines a statistical sampling of inelastic collisional transitions with an adiabatic channel representation. More technically, within this method the scattering Hamiltonian is diagonalized excluding the kinetic energy term in the basis of angular functions for the collision. Upon diagonalization, a set of adiabatic potential curves is obtained as a function of the distance ($R$) between the two colliders for each value of total angular momentum ($J_{\text{tot}}$). Each curve defines an entrance channel characterized by an effective centrifugal barrier. A collision is assumed to occur when the kinetic energy exceeds this barrier, leading to an open channel. In this framework, inelastic transition probabilities are determined solely by the number of open channels at a given energy, which are assigned equal statistical weight. A major advantage of this approach is its computational efficiency. Because the adiabatic curves are independent of the collision energy, they need to be computed only once for each value of the total angular momentum. Moreover, the basis of accessible rotational levels can be selectively truncated by retaining only the adiabatic curves that provide the dominant contributions to the scattering matrix. Taken together, these factors lead to a substantial reduction in terms of computational cost, convergence time, and memory requirements with respect to full quantum calculations. In terms of accuracy, the SACM is particularly well suited for systems that form long-lived intermediate complexes, where energy redistribution is expected to be statistical.
Moreover, while quasi-classical trajectory methods tend to lose accuracy at low collision energies, SACM is most effective in this regime, making the two approaches complementary depending on the temperature range considered.
Owing to the nature of the CO and H$_2$O projectiles and their interactions with collision partners in cometary comae, the SACM has already shown encouraging results for several systems, typically reproducing thermalized rate coefficients within a factor of two of exact close-coupling (CC) results \cite{godard2025promising,tonolo2025collisional}.

Despite these successes, it remains unclear whether SACM could reliably reproduce the essential quantum and structural signatures that govern low–temperature collisions. 
Owing to its intrinsically statistical formulation, the method would, in principle, appear insensitive to such effects, as conventional statistical approaches assume complete energy redistribution within the intermediate complex and depend only on global densities of states. However, the adiabatic channel representation of SACM could retain geometry-dependent dynamical information and interaction anisotropy, thereby conferring an unexpected sensitivity to the detailed topography of the potential energy surface and enabling the reproduction of near energy-resonant transfer and structural features.
%Owing to its intrinsically statistical formulation, the method would, in principle, appear transparent to such effects. However, its foundation in the adiabatic channel representation may confer an unexpected sensitivity to the anisotropic components of the interaction potential, thereby reproducing the corresponding quantum resonances and interference effects.
To date, no computationally affordable approach has convincingly demonstrated the capability to describe low temperature quantum and structural features in collisions involving heavy projectiles. In this work, we address this challenge by systematically evaluating the SACM ability to capture such effects.

%This limitation generates significant uncertainty where quantum resonances and interference effects become prominent. 
A particularly sensitive benchmark in this respect is provided by comparative studies of collisional rate coefficients for molecular isomers with similar rotational structures. These systems are indeed scarcely discernible by pure statistical approaches, as their relative dynamics depends on the distinct contributions of the anisotropic terms of the interaction potentials. Consequently, these comparisons constitute a stringent test of the method's ability to resolve subtle structural features. We present here a comparative study of the HCN and HNC isomers colliding with CO, which is the dominant projectile in comets at large heliocentric distances, where low kinetic temperatures amplify such effects. These two isomers have long attracted considerable attention in molecular physics, as their collisional coefficients exhibited distinct propensity rules depending on both the isomeric form and collision partner \cite{varambhia2007electron,hernandez2017rotational}.
Previous studies have demonstrated that accurate collisional rate coefficients for HCN and HNC are crucial for reliable molecular abundance determinations and chemical diagnostics in a wide range of environments. In particular, Hays \emph{et al.} \cite{hays2022collisional} proved experimentally that the isomeric differences between HCN and HNC have a strong impact on the collisional excitation in low temperature conditions,
influencing the inferred physical conditions and chemical compositions.

However, although both HCN and HNC isomers have been observed in numerous cometary comae %\cite{lis1997spectroscopic,irvine1997hnc,rodgers1998hnc,hirota1999observations,irvine2003hcn,lis2008hydrogen,biver2011molecular,villanueva2013modeling,agundez2014molecular,cordiner2014mapping,cordiner2017alma, cordiner2023gas},
\cite{irvine2003hcn,villanueva2013modeling,agundez2014molecular,cordiner2017alma,cordiner2023gas}, 
the only available collisional data of these species 
in such environments are the recently retrieved datasets limited to HCN in collision with CO and H$_2$O \cite{tonolo2025collisional,zoltowski2025collisional}.  
As a consequence, the wide variety of HNC/HCN abundance ratios (ranging from $\sim\,0.03$ to 0.3) in these environments remains puzzling, and the formation mechanism of HNC is still unclear \cite{mumma2011chemical,hays2022collisional}. In order to expand the collisional information for these systems, and test the reliability of the SACM in capturing prominent quantum effects with respect to isomeric variations, we decided to extend the existing dataset of HCN colliding with CO by deriving the corresponding rate coefficients also for its isomeric counterpart, HNC.

The computational protocol adopted to derive the (de)-excitation rate coefficients of HNC in collision with CO closely follows that previously established for the HCN and CO system \cite{tonolo2025collisional}. In the following, the main steps are briefly summarized, and the results are reported and discussed. For a more detailed description of the computational methods and convergence tests, the reader is referred to the Computational Methods section.
A global four dimensional interaction potential of the HNC and CO collisional system was constructed at the same level of accuracy of the potential for the HCN and CO system \cite{tonolo2025collisional}, based on coupled-cluster electronic structure theory, and subsequently scattering calculations were performed within the SACM statistical approach. 
As in the HCN–CO system, HNC forms strongly hydrogen-bonded, nearly collinear complexes with CO, with the H-atom directed toward either end of the CO molecule, preferentially the C end. The potential wells for HNC are significantly deeper than for HCN, reflecting the stronger N–-H bond compared to that of C–-H for HCN.

Using the HNC--CO interaction potential, the (de)-excitation cross sections were computed for transitions among the first 10 rotational levels of the HNC target molecule ($j_2$) and the first 9 levels of the projectile CO ($j_1$), accounting for more than 99\% of the population at 50\,K. These data were used to derive state-to-state and thermalized rate coefficients over the $5$--$50$\,K temperature range. The thermalized coefficients are commonly used in radiative transfer models of cometary comae under the assumption of thermal equilibrium between the rotational temperature of the projectile and the kinetic temperature of the surrounding gas. They were obtained by averaging the state-to-state rate coefficients over the CO rotational distribution (Eqs.~2 and 3 in Tonolo \emph{et al.} \cite{tonolo2025collisional}). 

\begin{figure*}[t]
  \begin{center}
  \includegraphics[scale=0.14]{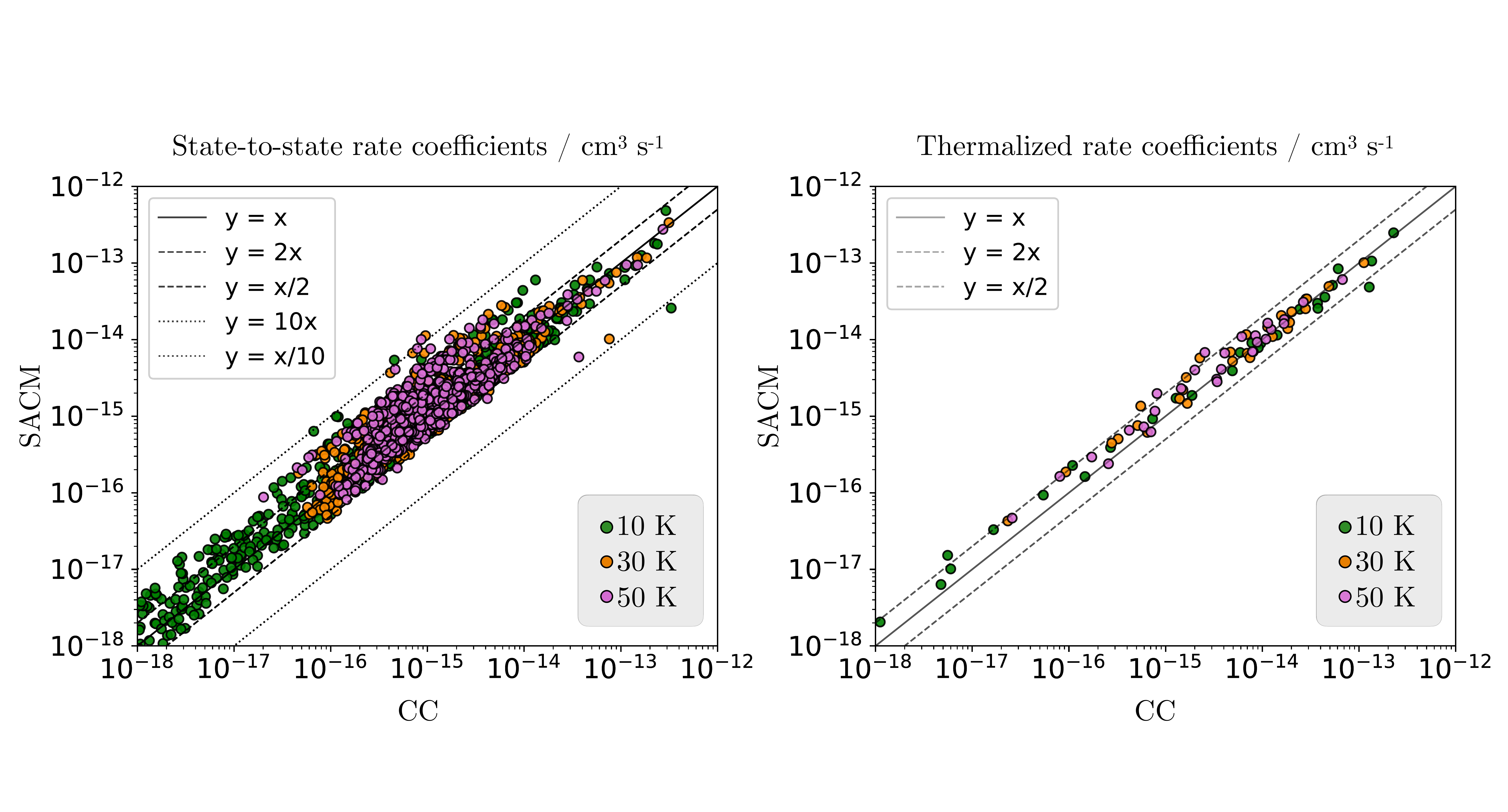}
  \caption{Comparison between the state-to-state (left panel) and thermalized (right panel) rate coefficients for the HNC and CO system, computed with the CC and SACM approaches at $J_{\text{tot}}=0$ and for $T = 10, 30$ and 50\,K.}
  \label{fig1}
  \end{center}
\end{figure*}
As was done for the HCN and CO system, the accuracy of the SACM in describing the HNC and CO collisional behavior was assessed through comparison with full-quantum CC calculations restricted to a single partial wave, $J_{\mathrm{tot}}=0$. The results for $T=10,30,$ and 50\,K are illustrated in Figure~\ref{fig1}. In accordance with the findings of Tonolo \emph{et al.} \cite{tonolo2025collisional}, the SACM showed good accuracy also for the HNC and CO collisional system. It reproduces state-to-state rate coefficients within a factor of two, and with maximum deviations within one order of magnitude. After thermal averaging, discrepancies are further reduced, yielding deviations within a factor of two for all the thermalized rate coefficients.
This was somewhat expected, since the wells in the HNC--CO interaction potential are nearly twice as deep as those of HCN--CO, favoring even more a statistical redistribution of the energy.

\begin{figure*}[t]
  \begin{center}
  \includegraphics[scale=0.19]{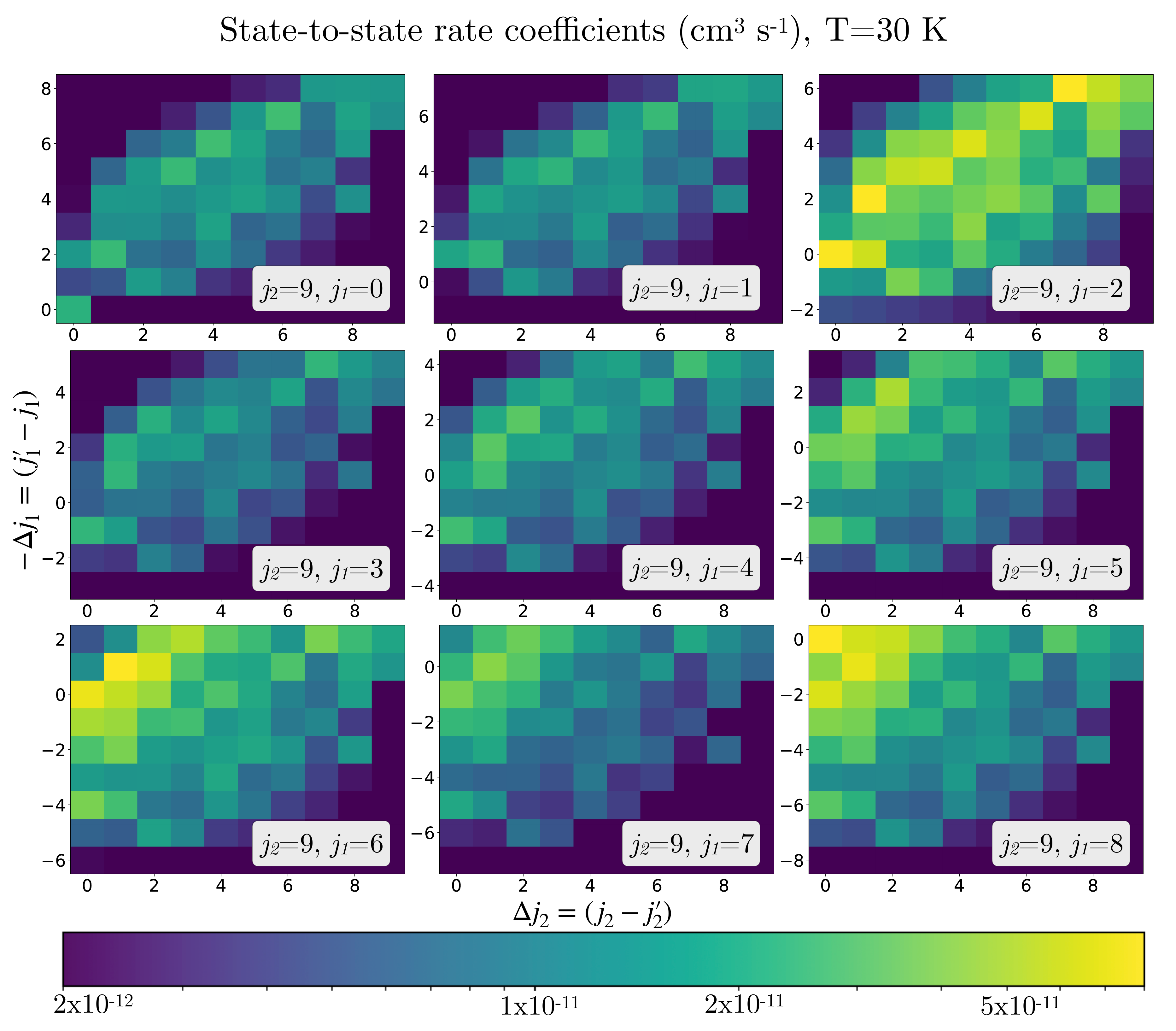}
  \caption{Variation of some state-to-state rate coefficients of the HNC and CO collisional system as a function of $\Delta j_2$ (x axis) and $-\Delta j_1$ (y axis), starting from $j_2=9$ and $j_1=0$--$8$ and for $T= 30$\,K. $\Delta j$ is defined as the difference between the initial ($j$) and final ($j'$) state.}
  \label{fig2}
  \end{center}
\end{figure*}
While this agreement demonstrates the reliability of SACM in reproducing overall collisional dynamics, it is not informative of its accuracy in describing quantum effects.
However, as illustrated in Figure~\ref{fig2} for the HNC and CO case and in Figure~7 of Tonolo \emph{et al.} \cite{tonolo2025collisional} for its isomeric counterpart, the SACM outcomes reveal clear signatures of near-resonant energy transfer regimes for both the HCN/HNC and CO systems. These purely quantum phenomena arise when the rotational energy levels of the colliding partners align closely and promote transitions that conserve the total internal energy, leading to dominant probabilities when $\Delta j_2 = -\Delta j_1$, $\Delta j_2 = -\Delta j_1 \pm 1$, and $\Delta j_2 = -\Delta j_1 \pm 2$.
A possible explanation for this behavior is that, as the collision energy increases, the number of accessible adiabatic channels grows, effectively increasing the density of energetically allowed final states connected to a given initial state. This enhanced channel density increases the likelihood of transitions into energetically nearby states, thereby favoring near-resonant energy transfer. In particular, at a given collision energy, transitions involving the smallest energy mismatch among the accessible channels tend to be favored, leading to an effective propensity toward quasi-resonant transitions.
However, since the SACM framework cannot resolve the resonances of bound and quasi-bound states, such features may be underestimated with respect to CC calculations. A comparison with full quantum results, including $J_{\text{tot}}>0$ values, would be highly valuable in this regard. However, the computational cost and memory requirements associated with such calculations, which involve more than 15000 scattering channels, makes them unaffordable for the present collisional systems. Further investigations on lighter systems would therefore be of interest to assess the accuracy of SACM in describing these effects.
%These results provided a first proof of the efficiency of the SACM method to preserve the effect of quantum features in its statistical adiabatic framework.  

\begin{figure*}[t]
  \begin{center}
  \includegraphics[scale=0.15]{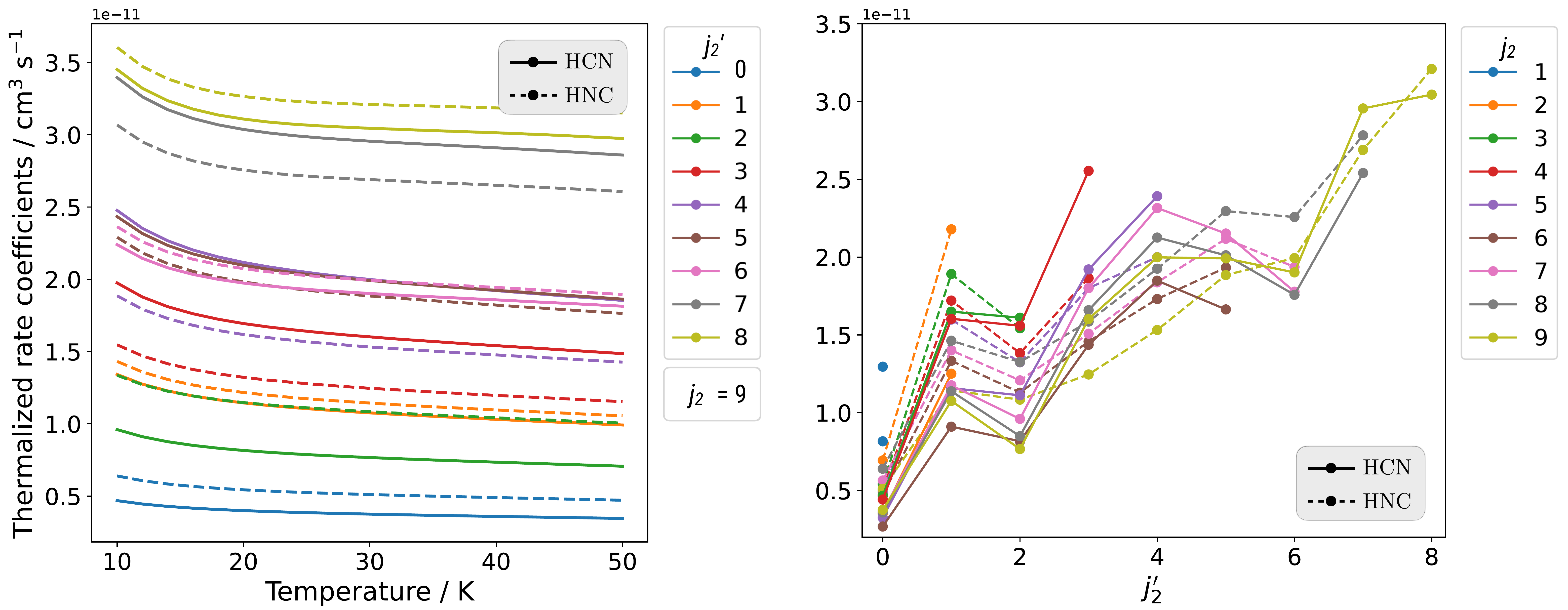}
  \caption{Dependence on the temperature and final rotational state of two sets of de-excitation thermalized rate coefficients for the $j_2 \rightarrow j_2'$ transitions of HCN (solid lines) and HNC (dashed lines) by collisions with CO. Left panel: variation with the temperature of the de-excitation rate coefficients starting from $j_2=9$. Right panel: variation of the rate coefficients as a function of $j_2'$, starting from $j_2=1$--$9$ and for $T = 30$\,K.}
  \label{fig3}
  \end{center}
\end{figure*}

\begin{figure*}[tb]
  \begin{center}
  \includegraphics[scale=0.15]{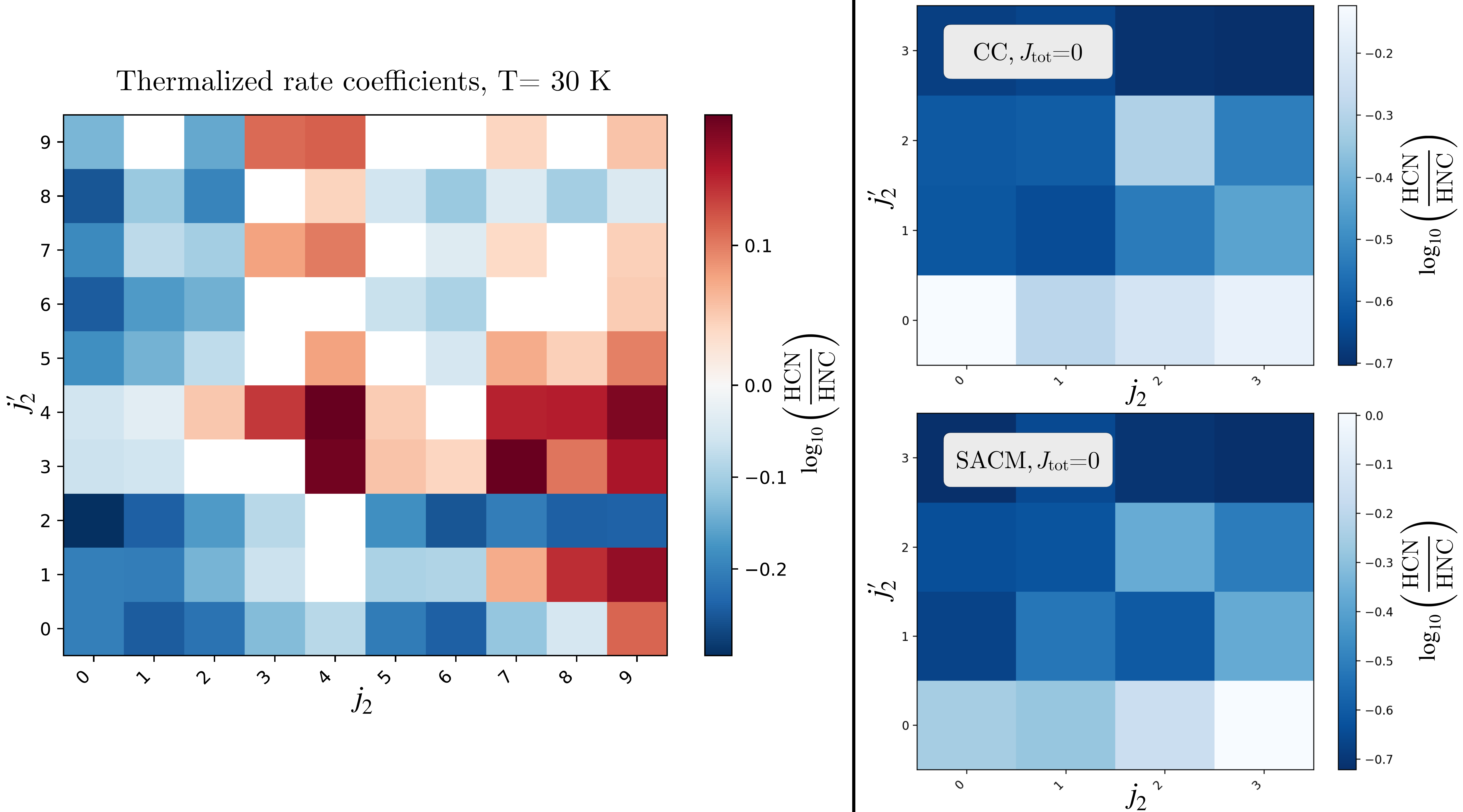}
  \caption{Left panel: Variation of the intensity ratio of the thermalized rate coefficients for the $j_2 \rightarrow j_2'$ transitions of HCN with respect to HNC in collision with CO, at $T= 30$\,K. Red squares indicate the transitions where collisions with HCN are favored, while blue squares where the (de)-excitation of HNC is promoted. Right panel: Variation of the intensity ratio of a reduced set of thermalized rate coefficients for the transitions involving the first 4 rotational levels of each target molecule, calculated with the full quantum CC method (upper plot) and the SACM approach (bottom plot), both restricted at the lowest total angular momentum ($J_{\mathrm{tot}}=0$).}
  \label{fig4}
  \end{center}
\end{figure*}

To probe the ability of the SACM to capture structural effects, we investigated its capability to discriminate isomer-dependent variations of the interaction potential in scattering calculations. In particular, we analyzed how structural differences between HCN and HNC are reflected in their thermalized collisional rate coefficients. Figure~\ref{fig3} presents a representative subset of thermalized de-excitation rate coefficients as a function of the temperature (left panel) and the final rotational state (right panel). In both panels, the de-excitation between the $j_2 \rightarrow j_2'$ levels of both HCN (solid lines) and HNC (dashed lines) are shown. The left panel reports the temperature dependence of transitions originating from $j_2 = 9$. A weak sensitivity to the kinetic temperature stands out: although absolute values are slightly enhanced at lower temperatures, the relative trends remain largely preserved across the explored thermal range. 
The right panel displays the dependence of the de-excitation rates on the final rotational level $j_2'$. In general, thermalized rate coefficients decrease with increasing $\Delta j_2$, reflecting the growing energy gap involved in the transition. However, no clear propensity rule emerges with respect to either the initial or final rotational states.
A distinctive feature common to both panels is the overall enhancement of the collisional de-excitation of HNC relative to HCN. This trend indicates that SACM is sensitive to differences in the depth of the interaction potential and their impact on collisional efficiency. The deeper potential well of the HNC and CO system indeed favors the formation of a longer-lived intermediate complex, which in turn enhances energy transfer during collisions.
To further quantify this effect, we examined the ratio of the thermalized rate coefficients of HCN and those for HNC for each transition. The left panel of Figure~\ref{fig4} displays these ratios at $T = 30$\,K, which are depicted in red when HCN exhibits larger rates and in blue when HNC dominates. As expected, transitions involving HNC are generally favored, particularly for low rotational levels, which are more sensitive to the differences in the potential well depth between the two isomers.
At higher values of $j_2$, rotational levels become progressively less influenced by variations in potential stability. In this regime, collisional (de)-excitation is more favored for the more stable isomer, HCN, leading to a gradual inversion of the dominant trend. Exceptions to this trend appear when the initial and/or final states correspond to quantum numbers 3 or 4, which seem to favor collisions with HCN. This behavior reflects the apparent favor of HCN transitions involving these states, as already shown in Figure~\ref{fig3}, that is not observed for HNC. At present, no clear physical explanation for this behavior has been identified.

These results demonstrate the high sensitivity of SACM to variations in the anisotropic components of the interaction potential associated with the two structural isomers. To further assess the robustness of this approach, we compared the HCN/HNC ratios of the rate coefficients obtained from SACM with those derived from full quantum CC calculations restricted to the lowest partial wave, $J_{\mathrm{tot}} = 0$.
The right panel of Figure~\ref{fig4} presents this comparison for transitions involving the lowest-energy rotational levels, which lie in the regime of maximum sensitivity to the potential anisotropy. An excellent agreement is observed between the two methods, with relative variations that are fully consistent across all examined transitions. This result provides an independent and stringent validation of the ability of SACM to capture isomer-specific effects in molecular collisions.

Taken together, our findings establish the SACM as a reliable and computationally efficient framework for describing low-temperature collision dynamics involving heavy projectiles. In this regime, fully quantum treatments become prohibitively expensive due to the rapid growth of scattering channels, severely limiting their applicability.
By contrast, the SACM method preserves the essential quantum and structural features governing energy transfer while achieving a substantial reduction in computational cost. 
In particular, it exhibited clear signatures of near-resonant energy transfer on the state-to-state rate coefficients of HCN and HNC colliding with CO, a hallmark of low-temperature quantum dynamics. Moreover, the isomeric ratios of the corresponding thermalized rate coefficients provided a sensitive probe of the different anisotropies of the interaction potentials.
%By contrast, the SACM method preserves the essential quantum features governing energy transfer while achieving a substantial reduction in computational cost.
%In particular, the state-to-state rate coefficients for HCN and HNC colliding with CO obtained with SACM exhibited clear signatures of near-resonant energy transfer, a hallmark of low-temperature quantum dynamics. Moreover, the isomeric ratios of the corresponding thermalized rate coefficients provided a sensitive probe of the different anisotropies of the interaction potentials. 
Finally, this work reports the first comprehensive dataset of state-to-state and thermalized rate coefficients for HNC in collision with CO, involving the 10 lowest rotational levels of HNC and the 9 lowest levels of CO over the $5$--$50$\,K temperature range. These data will support the interpretation of current and future observations of HNC in cometary comae, providing new insights into its formation pathways and helping to constrain the pronounced variations in isomeric abundance observed in these environments.

\section*{Computational Methods}
\label{sec:comp}
\begin{figure}[h!]
 \includegraphics[width=0.4\textwidth]{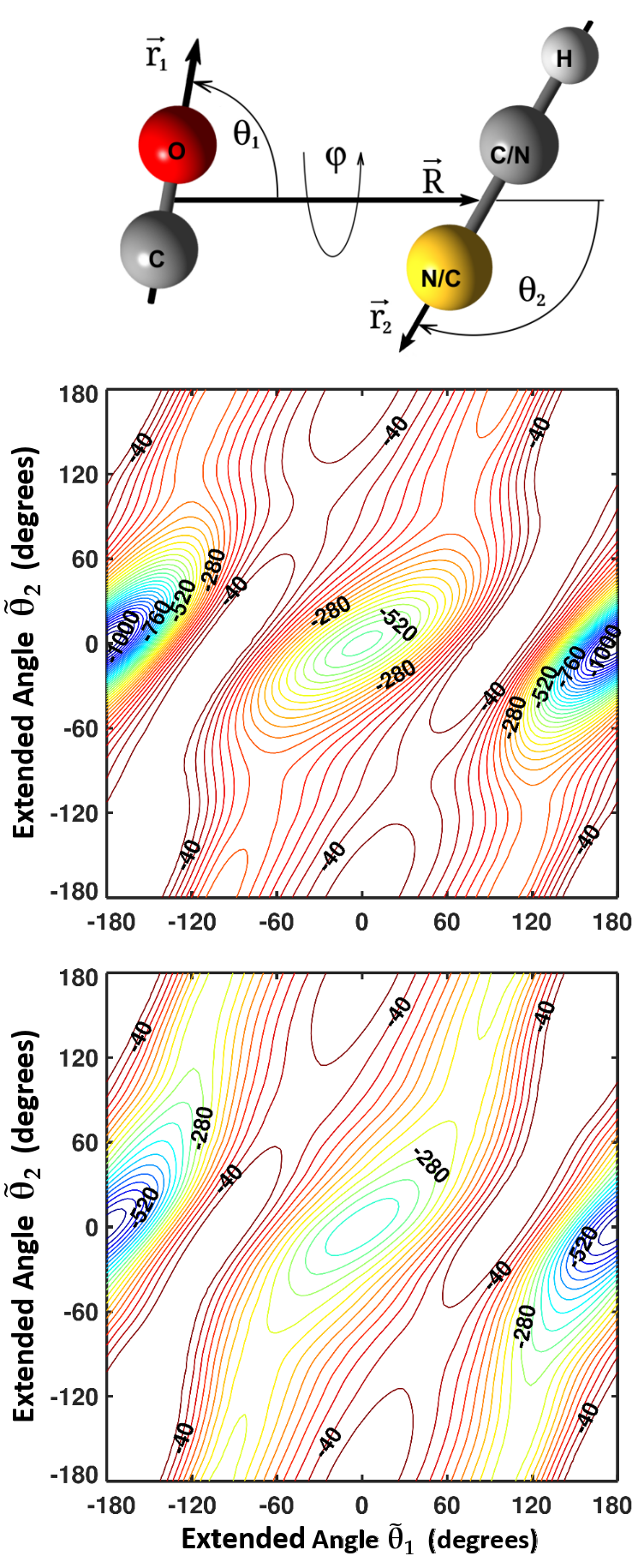}
 \caption{ (upper) Coordinates used to describe the CO--HNC/HCN interaction. (middle and lower) $R$-optimized contour plot of the CO--HNC (middle) and CO--HCN (lower) PESs as a function of the extended angles $\tilde{\theta}_1$ and $\tilde{\theta}_2$ for planar configurations ($\varphi=0^{\circ}$ and $\varphi=180^{\circ}$). For each pair of angles, the energy (given in cm$^{-1}$) is optimized with respect to the center-of-mass distance $R$. See text for details.}
  \label{SR}
\end{figure}

As shown in Fig.~\ref{SR}, the four-dimensional CO–HNC and CO–HCN PESs are described using Jacobi coordinates $R$, $\theta_1$, $\theta_2$, and $\varphi$. CO is monomer 1, while HNC/HCN is monomer 2. Here, $\vec{R}$ connects the centers of mass, and $\vec{r}_1$, $\vec{r}_2$ lie along the molecular axes. $R$ is the length of $\vec{R}$, $\theta_1$ and $\theta_2$ are the angles between $\vec{R}$ and $\vec{r}_1$, $\vec{r}_2$, and $\varphi$ is the dihedral angle between $\vec{R}\times\vec{r}_1$ and $\vec{R}\times\vec{r}_2$.

%As depicted in Figure~\ref{SR}, the coordinates used to represent the four-dimensional CO--HNC and CO--HCN Potential Energy Surfaces (PESs) are the Jacobi coordinates $R$, $\theta_1$, $\theta_2$, and $\varphi$.
%Note that for the representation of the PESs, CO is monomer 1, and HNC or HCN is monomer 2. 
%$\vec{R}$ is the vector between the centers of mass of the two fragments, and $\vec{r}_1$ and $\vec{r}_2$ are vectors aligned with each molecule.
%Coordinate $R$ is the length of vector $\vec{R}$, while coordinates $\theta_1$ and $\theta_2$ represent (respectively) the angles between $\vec{R}$ and the vectors $\vec{r}_1$ and $\vec{r}_2$. $\varphi$ is the dihedral (out of plane) torsional angle (angle between the vectors $\vec{R}\times\vec{r}_1$ and $\vec{R}\times\vec{r}_2$). 
%Notice that for $\theta_1=\theta_2=0^{\circ}$ the molecules are aligned, with the oxygen atom in the CO molecule pointing to the hydrogen atom in the HNC or HCN molecule.

\paragraph{Potential Energy Surface}

For construction of the PESs, both monomers were held rigid. 
For CO--HNC, the bond distances were fixed at $r_{\text CO}=1.128$~{\AA}, $r_{\text HN} = 0.994$~{\AA} and $r_{\text NC} = 2.163$~{\AA}.
%, consistent with the rotational constants of the molecules.
%Masses of $15.9949146221$, $12$, $1.007825032$, and $14.0030740052$~$u$ were used for $^{16}$O, $^{12}$C, $^{1}$H and $^{14}$N, respectively.

As was done in the past for other van der Waals linear dimers \cite{castro2019computational,quintas2020computational,gancewski2021fully,Desrousseaux2021,quintas2021theoretical,zadrozny2022,ajili2022theoretical,olejnik2023ab,bostan2024mixed},
the PES analytical representation was constructed using an automated interpolating moving least squares (IMLS) methodology, freely available as a software package under the name \textsc{autosurf} \cite{quintas2018autosurf}.
%As usual \cite{Dawes2018,majumder2016automated}, a local fit was expanded about each data point, and the final potential is obtained as the normalized weighted sum of the local fits.
The fitting basis and most other aspects of the IMLS procedure were similar as for other previous systems that have been described in detail elsewhere \cite{majumder2016automated,Dawes2018,quintas2018autosurf,quintas2021spectroscopy}.

%The shortest intermonomer center-of-mass distance represented is $R = 2.4$~{\AA} since all closer configurations are repulsive and beyond the energy range of interest. 
The short-range (high-energy) part of the PES was restricted by excluding regions with repulsive energies more than $6$~kcal/mol ($\sim2\,000$~cm$^{-1}$) above the separated monomers asymptote. 
The \textit{ab initio} data coverage in the fitted PES extended to $R=25$~{\AA}, while the zero of energy was set at infinite separation of the monomers.
%To guide the placement of high-level \textit{ab initio} data, a lower-level guide surface was constructed using $3\,500$ points at the explicitly correlated CCSD(T)-F12b/VDZ-F12 level \cite{Werner2011a}, distributed using a Sobol sequence \cite{sobol1976uniformly} biased to sample the short range region more densely. 
%Since the two fragments are not identical, and each has only C$_{\infty v}$ symmetry, there is no additional symmetry to respect/exploit in the PES beyond that of the angle $\varphi$, for which energies were only computed in the reduced angular range: $0<\varphi<180^{\circ}$.
All \textit{ab initio} calculations were performed using the Molpro electronic structure code package \cite{Werner2012-molpro} and using the explicitly correlated CCSD(T)-F12b method, where CCSD(T) denotes the coupled-cluster single and double approximation augmented by a perturbative treatment of triples, extrapolated to the complete basis set (CBS) limit. 
The CBS extrapolation was performed using the VTZ-F12 and VQZ-F12 bases \cite{peterson2008systematically} and the $l^{-3}$ formula \cite{feller2006sources}.
In total, $3\,776$ points were used, resulting in an estimated global root-mean-squared fitting error of $1.064 $~cm$^{-1}$ (excluding the long range where the fitting error is extremely small). The estimated error is $0.186 $~cm$^{-1}$ for geometries corresponding to energies in the wells (below the asymptotic zero).
To represent the long range (out to arbitrary distances), the PES switches smoothly to an analytic expression representing electrostatic, induction, and dispersion interactions between the fragments (truncated at 8th order). 
The long-range representation was produced using the recently released LRF software \cite{batista2026long}.
%The analytical representation of the PES is available from the authors upon request.

Figure~\ref{SR} shows a 2D comparative representation of the CO--HNC/HCN PESs (denoted $R$-optimized) as a function of the extended angles $\tilde{\theta}_1$ and $\tilde{\theta}_2$ for planar configurations.
The extended-angle coordinates have been described in detail elsewhere \cite{Dawes2013}.
%For planar geometries 
%($\varphi=0$ for quadrants II and %IV, 
%and $\varphi=\pi$ for quadrants I %and III),
%the plot describes the complete %ranges of  $\tilde{\theta}_1$ and %$\tilde{\theta}_2$, 
%relaxing the intermonomer distance %coordinate $r_0$ for each pair of %angles.
%This type of plot provides unique %insight into the isomers in the %system, since for many cases---those %(such as this one) without non-%planar minima---the plot will %include all isomers and any planar %isomerization paths between them, %making it easy to visualize planar %motions during which $\varphi$ %changes from $0$ to $\pi$.
Both systems are characterized by two minima: the global minimum (GM), a collinear arrangement where the C-atom of CO approaches the H-atom of HNC or HCN; 
and a local minimum (LM), also collinear, where the O-atom of CO approaches the H-atom of HNC or HCN. These collinear minima are seen across the middle of the two lower panels in Figure~\ref{SR}, with GM appearing twice in each plot due to the use of extended angles.
The energies and structural parameters are given in Table~\ref{table1}, where it is seen that for CO--HNC, GM and LM have well depths of $1031$~cm$^{-1}$ and $564$~cm$^{-1}$ respectively, whereas for CO--HCN, the corresponding well depths are substantially less at $604$~cm$^{-1}$ and $363$~cm$^{-1}$. %Nevertheless, these wells are all quite deep due the hydrogen bonding nature of the interactions, and the differences reflect the different hydrogen bonding interaction to an N-H bond rather than to a C-H bond. 
%As seen in Table ~\ref{table1}, for each system, the less stable LM structure has a significantly shorter separation between monomers (than the corresponding GM structure), partly due to the location of the CO monomer's center-of-mass. 

\begin{table}[t]
\caption{\label{table1} Geometric parameters and  potential energy for stable structures in the PES. Energies are given relative to the asymptote. Units are \AA\,, degrees, and cm$^{-1}$.}
	\centering
	\begin{tabular}{lrrrr}
     & \multicolumn{2}{c}{CO--HNC} & \multicolumn{2}{c}{CO-HCN} \\
	               & GM         & LM & GM         & LM       \\
	\hline
	$R$           & $4.400 $   & $ 4.150$  & $4.826 $   & $4.532 $   \\
	$\theta_1$ & $180.0$   & $0.0$   & $180.0 $   & $ 0.0$   \\
	$\theta_2$ & $0.0$    & $0.0$   & $0.0 $   & $0.0 $  \\
	$\varphi$        & ---     & ---  & ---   & ---     \\
	$V$           & $-1030.59 $ & $-563.98 $  & $-604.13$   & $-363.32 $ \\
    \hline
	\end{tabular}
\end{table} 

\paragraph{Scattering Calculations}
On top of the HNC and CO interaction potential, collision dynamics calculations were performed by using the HIBRIDON code \cite{alexander2023hibridon}.

%SACM calculations were performed by computing the adiabatic channels for each value of the total angular momentum, $J_{\text{tot}}$. 
The scattering matrices were constructed by including, at each total energy up to $\sim500$\,cm$^{-1}$, all open adiabatic channels, assuming equal probability for their occurrence. In this energy range, the highest accessible rotational levels are $j^{\text{max}}_2=18$ for HNC and $j^{\text{max}}_1=15$ for CO. However, employing the complete rotational basis resulted in substantial memory requirements, exceeding 15000 open channels already at $J_{\text{tot}}=2$. To reduce the computational cost, the basis was therefore truncated to  $j_2^{\text{max}}=12$ and $j_1^{\text{max}}=10$. The reliability of this truncation was assessed by verifying that, for $J_{\text{tot}}=2$, the convergence of the rate coefficients remained within 20\% of the values obtained with the full basis over the $5$--$50$\,K temperature range. %Since contributions from higher  $J_{\text{tot}}$ mainly increase the centrifugal barrier associated with collisional processes, the remaining deviations can reasonably be regarded as lying within the uncertainty of fully converged calculations. 
Using the reduced basis, a partial-wave expansion was carried out for $J_{\text{tot}}$ values from 0 to 180, ensuring convergence of the cross sections to better than 5\%. This procedure yielded state-to-state collisional cross sections for transitions involving rotational levels up to $j_2= 9$ for HNC and $j_1=8$ for CO. These cross sections were subsequently averaged over a Maxwell--Boltzmann energy distribution to obtain the corresponding state-to-state rate coefficients between 5 and 50\,K. 

To evaluate the accuracy of the SACM approach, a subset of rate coefficients was compared with results obtained from full quantum CC calculations.
The CC computations were performed over the $3$--$200$\,cm$^{-1}$ energy range, using very fine energy steps (0.2\,cm$^{-1}$) below 20\,cm$^{-1}$, which were gradually increased to 5\,cm$^{-1}$ above 150\,cm$^{-1}$. The rotational bases for HNC and CO were optimized by analyzing the effect of including different rotational levels on the convergence of inelastic cross sections, and selecting basis sets that ensured convergence within 2\%: between $2$ and $50$\,cm$^{-1}$, the first 26 rotational states of HNC and the first 24 states of CO were included, which were progressively increased to 30 and 28, respectively, at $200$\,cm$^{-1}$. In order to keep the computational cost manageable in the considered energy range, the angular part of the nuclear Schrödinger equation was restricted to a single partial wave with $J_{\text{tot}}=0$. The radial dynamics was then solved by numerical propagation over the $R=2.9$--$42.0$\,{\AA} range. CC state-to-state rate coefficients were calculated for temperatures between 5 and 50\,K, accounting for (de-)excitation processes among HNC--CO rotational levels below 80\,cm$^{-1}$. These results were finally compared with their SACM counterparts, which were obtained using the same rotational basis employed for data production, but considering only the adiabatic channels up to $200$\,cm$^{-1}$ and for $J_{\text{tot}}=0$. 

\section*{Data Availability Statement}
The data underlying this article will be made available through the EMAA \cite{faure2025excitation}, LAMDA \cite{schoier2005atomic} and BASECOL \cite{dubernet2024basecol2023} databases. The two PESs employed in this study (CO-HCN and CO-HNC) are provided as Supplementary Materials.

\section*{Authors Declarations}
The authors have no conflicts to disclose.

\section*{Supplementary Material}
The two potential energy surfaces (PESs) employed in this study (CO-HCN and CO-HNC) are provided as supplemental material. Within each potential, a Fortran program that enables the evaluation of the energy (in cm$^{-1}$) at any point on each potential as a function of the Jacobi coordinates is also present.

\section*{Acknowledgements}

The authors thank the Regional Council of Brittany and the ``Programme National de Planétologie'' (PNP) under the responsibility of INSU, CNRS (France) for supporting this work.
FL and FT acknowledge the support from the CEA/GENCI (Grand Equipement National de Calcul Intensif) for awarding access to the TGCC (Très Grand Centre de Calcul) Joliot Curie/IRENE supercomputer within the A0110413001 project.
ABP was supported by the Kummer Institute for Student Success of the Missouri University of Science and Technology.
RD and EQS are supported by the United States Department of Energy (DOE), Grant No.\,DE-SC0025420.

%\nocite{*}
\bibliography{aipsamp}% Produces the bibliography via BibTeX.

\end{document}